\documentclass[times,aps,prb,twocolumn,groupedaddress,showpacs]{revtex4-1}  
\newcommand{\width}{8cm}

\usepackage{graphicx} 
\usepackage{dcolumn} 
\usepackage{bm} 
\usepackage{times}
\usepackage{colordvi}

\begin{document}

\newcommand{\FIGURE}[2]{
       \typeout{FIGURE #1.eps}
       \begin{figure}[h]
       \includegraphics[width=\width]{#1.eps}
       \caption{#2}
       \label{#1}
       \end{figure}
}

\newcommand{\Fbox}[1]{\mbox{$\bigcirc\!\!\!\!${\scriptsize #1}$\;$}}


\title{From percolating to dense random stick networks: Conductivity model investigation}

\newcommand{\SCL}{Scientific Computing Laboratory, Institute of Physics Belgrade,
University of Belgrade, 11080 Belgrade, Serbia}

\author{Milan \v{Z}e\v{z}elj}
\email[Corresponding author: ]{milan.zezelj@ipb.ac.rs}
\affiliation{\SCL}
\author{Igor Stankovi\'{c}}
\affiliation{\SCL}

\begin{abstract}
In a Monte Carlo study the conductivity of two-dimensional random stick systems
is investigated from the percolation threshold up to ten times the percolation 
threshold density. We propose a model explicitly depending on the stick density 
and junction-to-stick conductance ratio. The model describes the transition from
the conductivity determined by the structure of a percolating cluster to the
conductivity of the dense random stick networks. The model is motivated by the
observed densities of the sticks and contacts involved in the current flow. The
finite-size scaling effects are also included in the description. The derived model
for conductivity should be broadly applicable to the random networks of the rodlike
particles.
\end{abstract}





\maketitle

\section{INTRODUCTION}
Recently there has been an increasing interest in the networks of randomly
distributed stick (rodlike) particles~\cite{2003_apl_Ramasubramaniam,
2005_apl_Bo, 2010_epl_Seppala, 2010_apl_Sangwan}, due to the development
in the area of conductive nanoparticles, such as the carbon nanotubes,
silicon, copper, and silver nanowires~\cite{2012_ns_Bergin}, and the promising
applications in electronics~\cite{2005_prl_Kumar}, optoelectronics~\cite{2004_nl_Hu},
and sensors~\cite{2008_nature_Cao}. The conventional electronic composites
containing stick particles as a filler can be used as a conducting channel
in the thin-film network configurations~\cite{2006_apl_Kumar, 2008_acsn_Engel,
2011_apl_Chandra}. Probably, the most important characteristics of the
stick networks, are the conductivity, either electric~\cite{2006_prb_Hu} or
thermal~\cite{2010_prl_Volkov}. The conductivity dependence on the stick
density and system geometry needs to be taken into account in any device
design~\cite{2007_prb_Behnam}. The percolation models~\cite{2003_ipt_Stauffer,
1983_prl_Balberg} are often used to model an onset of the high electrical
conductivity in the composites consisting of the conductive sticks in the
insulating matrices~\cite{2003_apl_Ramasubramaniam, 2004_nl_Hu, 2006_nl_Unalan}.

The percolation theory predicts that the electrical conductivity
of the composite materials with the conductive filler density $n$ above,
but close to the percolation threshold $n_c$, increases with the
density by a power scaling law $\sigma \sim (n - n_c)^t$, with the
universal conductivity exponent $t \approx 1.29$ for two-dimensional (2D)
systems~\cite{2003_ipt_Stauffer}. While the conductivity scaling
law is expected to be applicable only near the percolation threshold,
in many experiments the scaling law was used over a much larger range
of concentration, but with the nonuniversal values of the conductivity
exponent~\cite{2004_nl_Hu, 2006_nl_Unalan, 1985_pl_Noh}. Hu {\it et al.}~\cite{2004_nl_Hu}
obtained the nonuniversal value $1.5$ for the conductivity exponent using
the conductivity scaling law for fitting the experimental data for
ultrathin carbon nanotube networks operating from the percolation threshold
up to about ten times the percolation threshold density.
Several numerical studies confirmed the observed nonuniversality of the
conductivity exponent when the stick density is well above the percolation
threshold~\cite{2004_prb_Keblinski, 2010_prb_Hazama, 2010_pre_Li}. Keblinski
{\it et al.}~\cite{2004_prb_Keblinski} demonstrated that the universal power
law holds from the percolation threshold $n_c$, to about twice its value $2 n_c$.
For higher stick density, $n > 2n_c$, and in two limiting cases they
observed that the conductivity scaling exponent becomes: (i) slightly higher
than $1$ when the junctions are superconductive and only the stick conductance
is the limiting factor for the current flow through the system and (ii)
close to $1.75$ when the sticks are superconductive and the contact conductance
is the limiting factor. Li {\it et al.}~\cite{2010_pre_Li} showed that the
conductivity exponent significantly varies with the junction-to-stick conductance
ratio for lower stick densities up to $2 n_c$. The broad range applicability of
the conductivity scaling law was explained by the presence of the long-range
correlations in the distribution of the conductive sticks in the
system~\cite{2003_hm_Sahimi}. We will demonstrate that the nonuniversality of the
conductivity exponents is a consequence of a transition from the percolating
to dense stick networks.

In this paper, we numerically investigate the conductivity of the stick systems
from the percolation threshold up to ten times the percolation threshold density. 
We show that it is not appropriate to use a simple scaling law to describe the 
conductivity dependence on the density both for finite and dense systems. 
Based on the Monte Carlo simulation results, a model is proposed describing the 
conductivity dependence on the stick density and the different junction-to-stick 
conductance ratios. The proposed model is valid for the different stick-like 
nanoparticles (e.g., the carbon nanotubes and the nanowires). The model is 
motivated by the observed structural characteristic (i.e., the density of the 
total sticks and contacts involved in the current flow through the system). The 
finite-size effects, especially pronounced in the vicinity of the percolation 
threshold, are included in the generic description for the conductivity of stick 
systems.

\section{NUMERICAL METHOD}
Monte Carlo (MC) simulations are coupled with an efficient iterative algorithm
implemented on the grid platform and used to investigate the conductivity
of stick systems~\cite{2010_pre_Li, 2002_cpc_Stankovic, 2012_cpc_Balaz}.
We have considered the two-dimensional systems with isotropically placed
widthless sticks of length $l$. The centers of the sticks are randomly positioned 
and oriented inside the square system with size $L$. Two electrodes
(i.e., conducting bars) are placed at the left and right sides. The top and bottom 
boundaries of the system are free and nonconducting. The free boundary 
conditions are more consistent with the finite-size rodlike nanoparticle networks 
in practice~\cite{2004_nl_Hu, 2008_nature_Cao, 2006_apl_Kumar, 2008_acsn_Engel,
2011_apl_Chandra}. Two sticks lie in the same cluster if they intersect.
The system percolates (conduct) if the electrodes are connected with the same cluster.
The behavior of the stick percolation is studied in terms of the stick density
$n = N / (L / l)^2$, where $N$ is the total number of sticks and $L / l$ is the
normalized system size. The percolation threshold of the infinite system is
defined by the critical density $n_c \approx 5.63726$ (Refs.~\cite{2009_pre_Li,
2012_pre_Zezelj}). To evaluate the conductivity of the stick systems we introduce two
different conductances: (1) the conductance of the entire stick $G_s$ and (2) the
conductance due to the stick-to-stick junction $G_j$. We assume diffusive
electrical transport through the stick typical for the rodlike nanostructures
(carbon nanotubes and nanowires) whose length is larger than the mean
free path of the electrons~\cite{2008_jpcm_Zhou}. According to the diffusive
electrical transport the electrical resistance of a stick segment is
proportional to the length of the segment~\cite{2004_nl_Park}. In our
simulations, each stick-stick junction is modeled by an effective contact
conductance regardless of the type of the junction, following the simplified
approach of the authors of Refs.~\cite{2005_prl_Kumar, 2007_prb_Behnam, 2010_pre_Li}.
Therefore, if two sticks intersect a junction with the fixed conductance
$G_j$ is created at the intersection point. If a stick intersects an electrode
the potential of the electrode is applied to the intersection point.
Kirchhoff's current law was used to balance the current flow through each
node of the created network. An iterative equation solver (i.e., conjugate
gradient method with Jacobi preconditioner) has been employed to solve a
large system of the linear equations following from the current law~\cite{2010_pre_Li,
1994_Shewchuk}. After obtaining the total current $I$ under an applied
voltage $V$ the macroscopic electrical conductivity of the system is evaluated
as $\sigma = I / V$ (Ref.~\cite{my_1}). Monte Carlo simulations have been
performed for a wide range (i.e., $G_j/G_s = 0.001$ to $1000$) of
junction-to-stick conductance ratios (cf. Refs.~\cite{2000_science_Fuhrer,
2001_prb_Buldum, 2000_jpcb_Yu, 2010_nn_Franklin, 2006_prb_Bid, my_2}). Finally,
for each set of the system parameters, the electrical conductivity is averaged
over the $N_{\rm MC}$ independent MC realizations. To obtain the same
precision for the finite-size systems $N_{\rm MC} = 64000$ realizations are used
for the systems with normalized size $L/l = 10$ down to $N_{\rm MC} = 4000$ for the
largest system $L/l = 40$ studied. Using the appropriate functions for the fitting
data and the least-squares fitting methodology~\cite{2012_pre_Zezelj}, good fits
with high correlation factors ($R^2 > 0.998$) were obtained for all analyzed systems.

\section{RESULTS AND DISCUSSION}
As already mentioned, the numerical estimates of the conductivity exponent $t$ are
based on the linear fit of the MC results for the logarithms of the conductivity
$\sigma$ and density $n - n_c$ (Refs~\cite{2003_apl_Ramasubramaniam, 2004_prb_Keblinski,
2010_prb_Hazama, 2010_pre_Li}). The estimates therefore rely on the assumption
that $\sigma$ obeys the simple power-law dependence over a quite extended density range.
As there exists no justification of such an assumption, we have investigated
in detail the sbehavior of the conductivity $\sigma$ as we move away from the
critical point. A local (density dependent) conductivity exponent is defined as
$t(n)$ by~\cite{1978_prb_Bernasconi, 2006_prl_Grimaldi}
\begin{equation}
t(n) = \frac{n - n_c}{\sigma} \frac{d \sigma}{d n}. \label{t_n}
\end{equation}
The dependence of the local conductivity exponent $t(n)$ on the stick density
$n$ and the ratio of the stick-stick junction conductance ($G_j$) to stick conductance
($G_s$) (i.e., $G_j/G_s$) is shown in Fig.~\ref{figure_1}. As one can see from a
coarse observation, when the stick density approaches the percolation threshold
$n_c$ from above the local conductivity exponent converges to the universal value
for 2D systems $t(n_c) \approx 1.29$ for all $G_j/G_s$ values. The fine behavior
of the local conductivity exponent for finite-size systems in the vicinity of the
percolation threshold will be discussed later in this section. With the increasing
concentration $n$, the local conductivity exponents $t(n)$ change quickly from the
universal value $t(n_c)$, taking the values in a wide range $1 \leq t(n) \leq 2$.
From Fig.~\ref{figure_1}, one can see that the local conductivity exponents $t(n)$
for the conductance ratio higher than 2 ($G_j/G_s>2$) is a monotonically decreasing
function of the stick density $n$ which converges to 1 from above. Somewhat surprisingly,
for the conductance ratios lower than 1 (i.e., $G_j/G_s<1$), the local exponent $t(n)$
is not a monotonic function and has a local maximum. The observed density where the local
conductivity exponent reaches a maximum is decreasing with the conductance ratio $G_j/G_s$.

\FIGURE{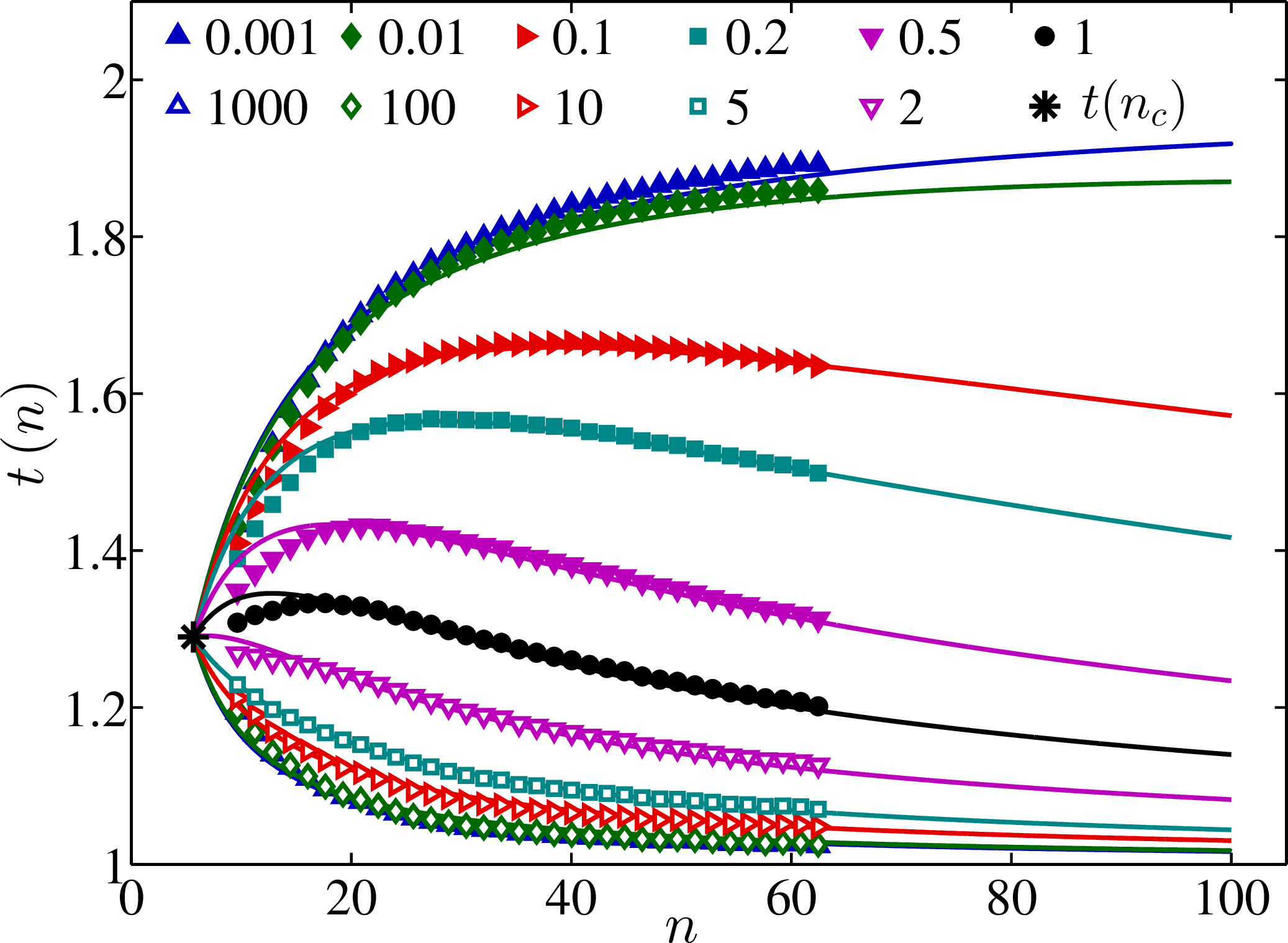}{(Color online) The dependence of the local conductivity
exponent $t(n)$ on the stick density $n$ and junction-to-stick conductance
ratio $G_j/G_s$. The points are MC simulation results obtained using 
Eq.~(\ref{t_n}) for the system size $L/l=20$. The values are given for
the conductance ratios $G_j/G_s = 0.001, 0.01, 0.1, 0.2, 0.5, 1$ (filled),
and their inverse values $1000, 100, 10, 5, 2$ (transparent). The error bars
are smaller than the size of the points. The star marker denotes the expected
universal value for the conductivity exponent at the percolation threshold $t(n_c)$.
The lines represent the local conductivity exponents $t(n)$ obtained from the
conductivity model for an infinite-size system, Eq.~(\ref{cond_model_inf}).}

To explain the observed behavior of the exponent $t(n)$ at the higher densities
$n>2n_c$, one needs to look into the structure of the dense conducting stick systems.
Figure~\ref{figure_2} shows the densities of the sticks $n^I$ and junctions $n_j^I$
that carry the current through the system. For sufficiently high stick
densities ($n>2n_c$), one can see that almost all the sticks and junctions in the
system contribute to the conductivity and that the density of the current-carrying
junctions increases with the stick density $n$ by a square power law $n_j^I\sim n^2$.
The reason for this is that the mean number of contacts per stick is proportional to the
stick density, see Ref.~\cite{2011_nt_Heitz}. Also, for a sufficiently high stick
density $n$ the current-carrying stick density $n^I$ is proportional to $n$.
Therefore, when the stick density is well above the percolation threshold $n_c$,
the conductivity of the system can be modeled as an equivalent serial conductance
$n$ sticks in parallel and $n^2$ junctions in parallel
\begin{equation}
\sigma \sim \frac{1}{b n^{-1} / G_s + n^{-2} / G_j},\label{sigma_dense}
\end{equation}
where $b$ is a constant parameter. One can see that the square term $n^{-2}/G_j$,
originating from the junctions, converges faster to zero than the linear term $bn^{-1}/G_s$.
This explains the conductivity exponent $t(n)$ approaching to 1 when the stick
density is sufficiently high (i.e., $n \gg G_s / G_j$) and the existence of the local
exponent maximum in Fig.~\ref{figure_1}. If the sticks are much more conductive 
than the junctions (e.g., $G_j/G_s=0.01$) the density where the local conductivity exponent
starts to converge to 1 is high and computationally unreachable in the MC simulations
shown in Fig.~\ref{figure_1}. Only in the limiting case when the sticks are superconductive
and the conductance ratio approaches zero, i.e., $G_j/G_s \to 0$, should the
conductivity exponent $t(n)$ converge to 2 with the increasing density $n$, which is
consistent with Keblinski {\it et al.}~\cite{2004_prb_Keblinski}. In the other limit,
when the junctions are superconductive (i.e., $G_j/G_s \to \infty$) the local
conductivity exponent $t(n)$ should have the fastest convergence to 1.

\FIGURE{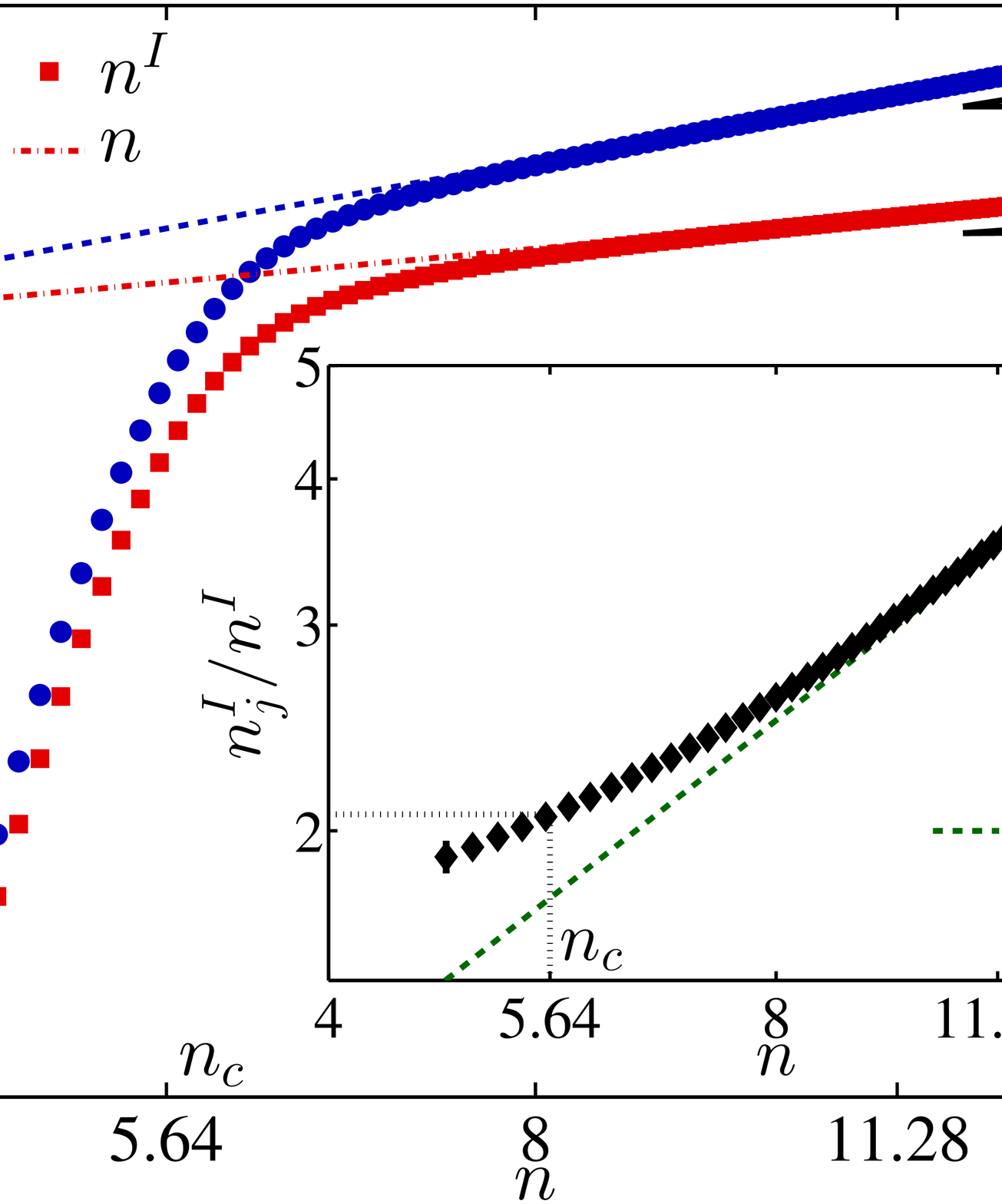}{(Color online) The density of junctions $n_j^I$ and sticks
$n^I$ involved in the current flow through the system is compared with
the density of all junctions $n_j$ and sticks $n$ in the system of size
$L/ l=20$. For higher stick densities $n$ almost all junctions and sticks
will carry some current. The error bars are smaller than the size of the
points. Inset: The density ratio of the current-carrying junctions to
current-carrying sticks $n_j^I/n^I$ is higher than the density
ratio of all junctions to all sticks $n_j/n$ in the system. At the
percolation threshold this ratio is about 2 [i.e., $n_j^I/n^I=2.0(1)$].}

At the densities close to the percolation threshold $n_c$, only a fraction but not all the
sticks and junctions in the system contribute to the conductivity, by carrying some
current. From Fig.~\ref{figure_2} (inset), one can see that at the percolation threshold
$n_c$, the density of the current-carrying junctions is about two times higher than the density
of the current-carrying sticks [i.e, $n_j^I/n^I=2.0(1)$]. From the framework of the percolation
theory we cannot determine a density-dependent factor of proportionality in the conductivity
power law [i.e., $\sigma \sim (n - n_c)^t$]. Instead, we fit the factor of proportionality
with an expression for the dense systems [i.e., Eq.~(\ref{sigma_dense})] and obtain
$1 / \left[b n^{t-1}/G_s + (n + n_c)^{t - 2}/G_j\right]$. This relation explicitly includes the 
previous observation that there is almost exactly two times more current-carrying junctions 
than current-carrying sticks at the percolation threshold. For a general conductivity
description of the infinite-size systems we obtain
\begin{equation}
\sigma = a \frac{(n - n_c) ^ t}{b n^{t - 1} / G_s + (n + n_c)^{t - 2} / G_j}, \label{cond_model_inf}
\end{equation}
where $a = 0.027(1)$ and $b = 0.061(3)$ are fitting parameters calculated using the
least-squares fitting methods. The solid lines in Fig.~\ref{figure_1} denote the
local conductivity exponents $t(n)$ calculated from Eq.~(\ref{t_n}), using
the model for an infinite system given by Eq.~(\ref{cond_model_inf}), for a wide
range of conductance ratios $G_j/G_s=0.001$ to $1000$. Deviations between the modeled
and MC values for local conductivity exponent $t(n)$ are comparable to the statistical
errors. 

\FIGURE{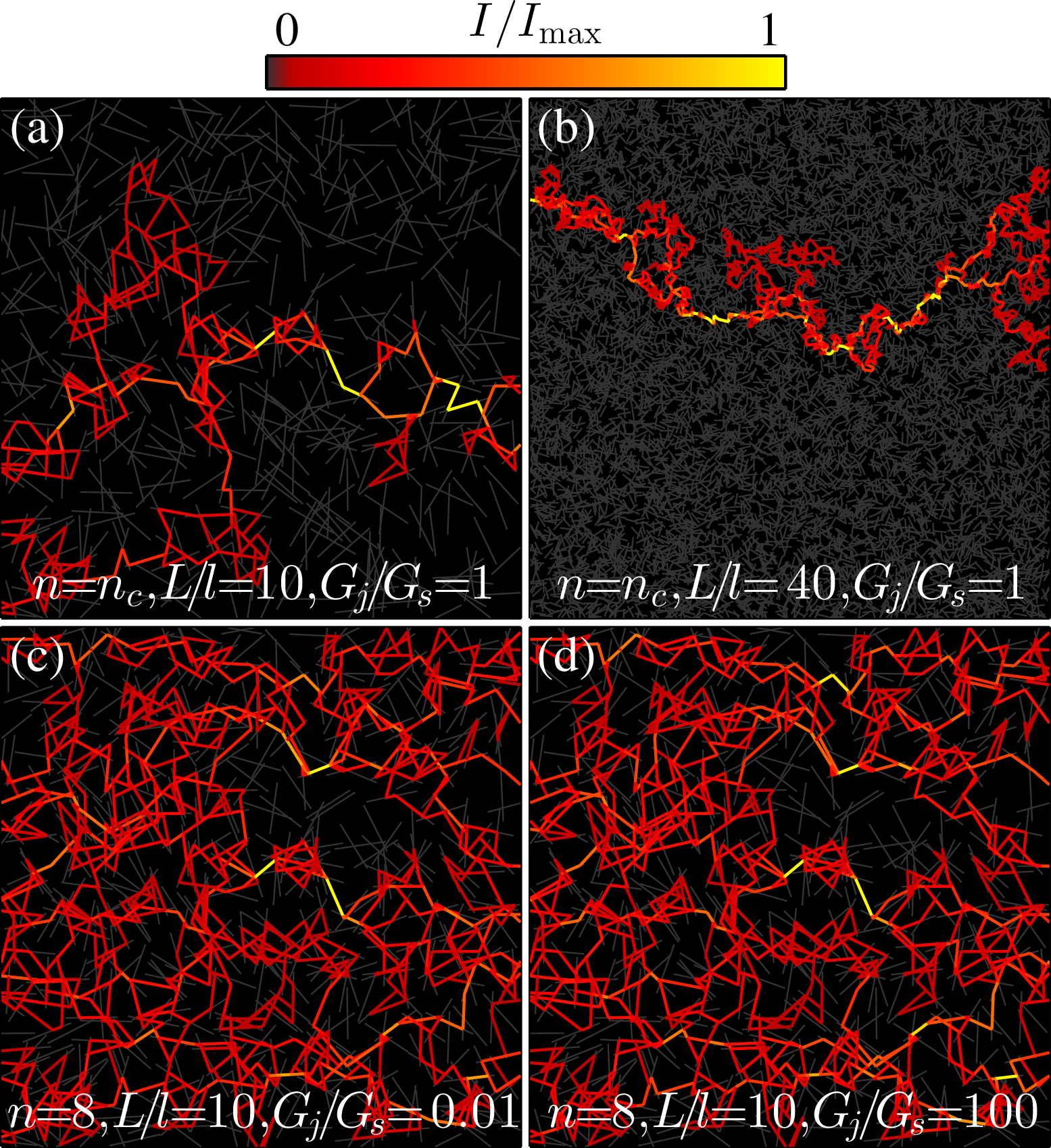}{(Color online) Simulated current (a) and (b) at different
system sizes and (c) and (d) junction-to-stick conductance ratios. The current
through a stick $I$ is given relative to the maximal current in the system
$I_{\rm max}$. There is a large difference in the fraction of the system
involved in the current flow between the two nominally identical films in term of
density ($n=n_c$) and junction-to-stick conductance ratio ($G_j/G_s=1$) for two
different system sizes $L/l=10$ and $40$. The current redistribution with the
increasing junction-to-stick conductance ratio $G_j/G_s$ is visible from (c) and (d).
(c) If junctions are weakly conductive (i.e., $G_j/G_s=0.01$) the maximal current
is flows along the shortest path with the least junctions. (d) For high junction
conductance values (i.e. $G_j/G_s=100$), the total current is evenly carried by the
larger number of shortest paths connecting electrodes. This effect is only visible at
higher densities (e.g., $n=8$) where several paths connecting electrodes exist.} 

\FIGURE{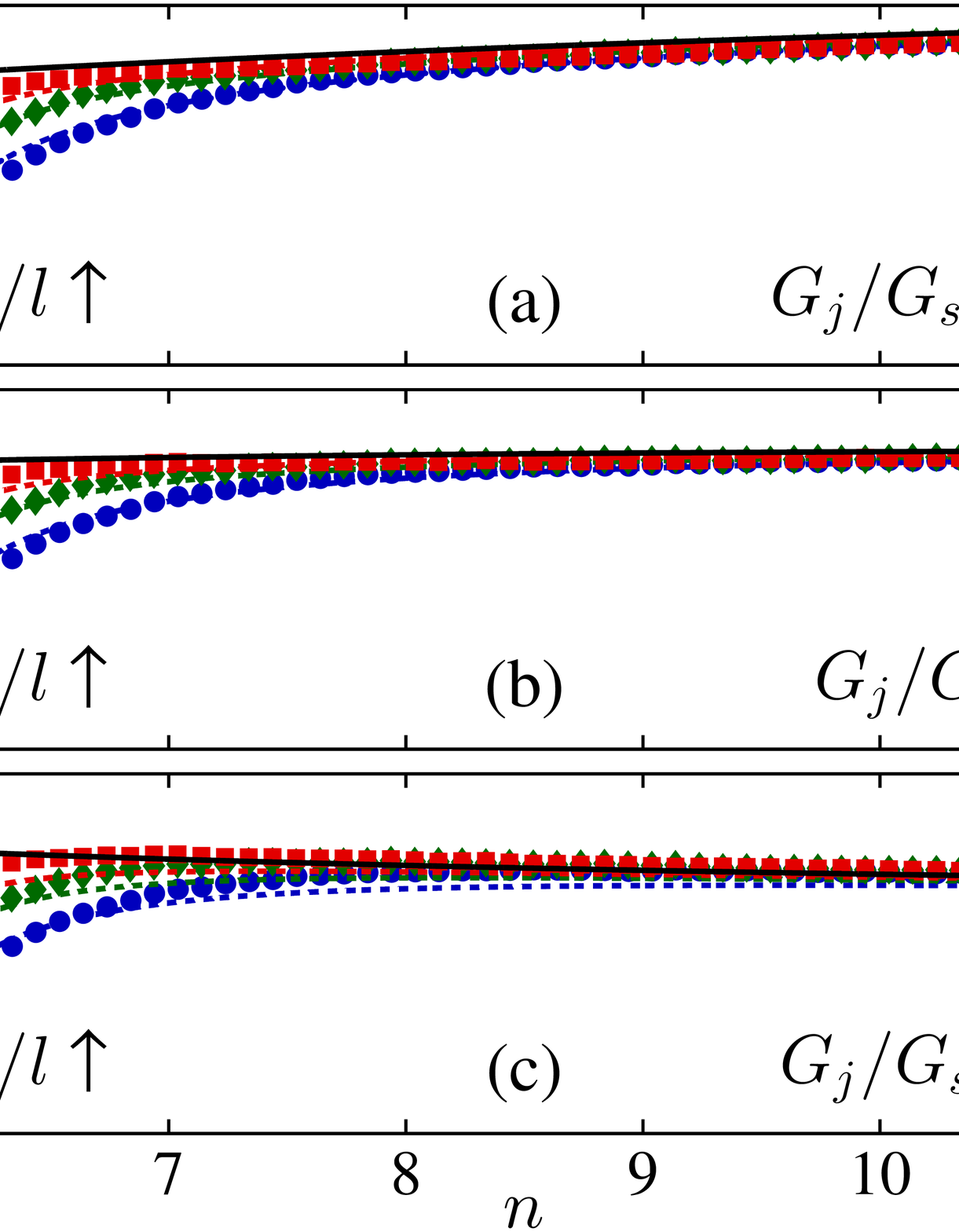}{(Color online) The local conductivity exponents $t(n)$ for
the stick systems with increasing size $L/l = 10, 20$, and $40$ and for
three conductance ratio values (a) $G_j/G_s = 0.01$, (b) $1$, and (c) $100$.
The direction of the increase of $L/l$ is indicated on the graphs. The points
are obtained from the MC simulations and calculated using Eq.~(\ref{t_n}).
The error bars are smaller than the size of the points. The solid line
represents the local conductivity exponent $t(n)$ for the infinite system
obtained from Eq.~(\ref{cond_model_inf}), while the dashed lines denote the local
conductivity exponents $t(n)$ obtained from the model that includes finite-size effects,
Eq.~(\ref{cond_model_fin}). The star marker denotes the expected value for the
conductivity exponent of the infinite system at the percolation threshold $n_c$.}

\FIGURE{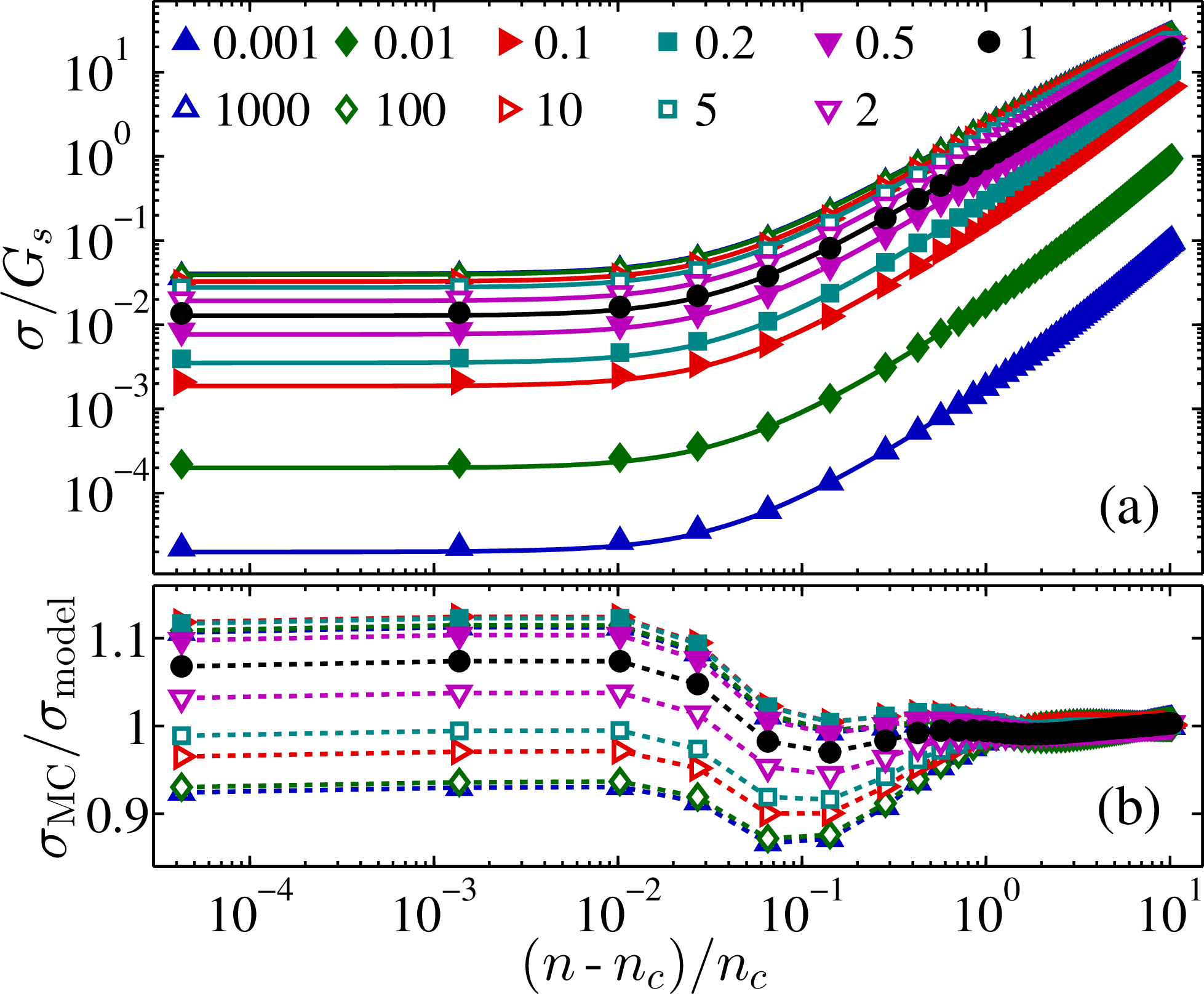}{(Color online) (a) Conductivity as a function of $(n - n_c) / n_c$
is obtained from the MC simulations (points) for the stick system of size $L/l = 20$
and the junction-to-stick conductance ratio from $G_j/G_s=0.001$ to $1000$ (from
bottom to top). The lines denote values obtained from the conductivity model for
the finite-size systems given by Eq.~(\ref{cond_model_fin}). (b) The conductivity
ratio between the MC simulation results $\sigma_{\rm MC}$ and the values obtained
from the model $\sigma_{\rm model}$ for corresponding finite-size systems,
Eq.~(\ref{cond_model_fin}). The error bars are smaller than the size of the points.}

Figure~\ref{figure_3} illustrates the structure of the percolating cluster
[Figs.~\ref{figure_3}(a) and \ref{figure_3}(b)] and the redistribution of the current
in the dense stick networks due to the junction-to-stick conductance ratio increase
[Figs.~\ref{figure_3}(c) and \ref{figure_3}(d)]. The current through a stick $I$ is
given relative to the maximal current in the system $I_{\rm max}$. As one can see from
Figs.~\ref{figure_3}(a) and \ref{figure_3}(b), the percolating cluster consists of 
a few sub-clusters connected by high current links. This explains why on average
more junctions than sticks are needed to shortcut the electrodes. For a large, 
but finite-size systems at the percolation threshold, the density of the current-carrying 
junctions decreases as $n_j^I \sim (L / l)^{-\beta/\nu}$ with normalized system size $L/l$, 
where $\beta = 5/36$ for 2D systems~\cite{2003_ipt_Stauffer}. Also, the density of 
the current-carrying sticks at the percolation threshold is $n^I\sim (L / l)^{-\beta/\nu}$. 
As a result the densities of the current-carrying sticks and junctions decrease with system 
size [cf. Figs.~\ref{figure_3}(a) ($L/l = 10$) and \ref{figure_3}(b) ($L/l = 40$)].
Furthermore, the density ratio $n_j^I/n^I$ at the percolation threshold converges to
a constant value with the increase of the system size, see Fig.~\ref{figure_2}. At higher
stick densities (i.e., $n = 8$) one can see that current flows along many parallel paths
connecting electrodes. An increase of the junction-to-stick conductance ratio $G_j/G_s$
results in the more uniform redistribution of the current [cf. Figs.~\ref{figure_3}(c) 
($G_j/G_s = 0.01$) and \ref{figure_3}(d) ($G_j/G_s = 100$)]. For weakly conductive junctions
(i.e., low  conductance ratio $G_j/G_s=0.01$), most of the current flows through a shortest
path with the least  junctions along. With the increase of the junction conductance
several parallel paths become visible. As a result, the total current through the system is
more evenly distributed, resulting in the higher conductivity. This is also expected from
Eq.~(\ref{cond_model_inf}).

If we compare the infinite system model prediction and MC simulation results in 
Fig.~\ref{figure_1} close to the percolation threshold, we observe a deviation
between the predicted and simulated values. This deviation is a result of the finite-size
effects, since the MC results in Fig.~\ref{figure_1} are calculated for the large 
but finite-size system (i.e, $L/ l = 20$). The convergence of the local conductivity 
exponents with the increasing system size is shown in Fig.~\ref{figure_4}. The points
are MC simulation results for the systems with sizes $L/l = 10, 20$, and $40$ and the 
solid line denotes the model for an infinite system given by Eq.~(\ref{cond_model_inf}). 
For the finite-size systems close to the percolation threshold we observe a large 
deviation of the local conductivity exponent $t(n)$ from the model. The local 
conductivity exponent decreases with the decreasing system size and can be even 
lower then 1 [$t(n)<1$]. This is result of a nonzero conductivity value
for the finite-size systems at the percolation threshold ~\cite{2003_ipt_Stauffer}.
Therefore, the model should be adapted for the finite-size systems. The
finite-size scaling arguments~\cite{2003_ipt_Stauffer, 2012_pre_Zezelj,1990_prb_Gingold}
suggest that the conductivity $\sigma$ depends on the system size $L$ as
\begin{equation}
\sigma \sim (n - n_c)^t f\left[\frac{\xi(n)}{L}\right], \label{sigma_L_l}
\end{equation}
where $\xi(n) \sim l |n - n_c|^{-\nu}$ is the correlation length that measures the linear
extent of the largest cluster. For 2D systems the correlation-length exponent is
$\nu=4/3$ (Ref.~\cite{2003_ipt_Stauffer}). For the infinite system above the percolation
threshold [i.e., $\xi(n) / L \to 0$] the conductivity follows the simple scaling
law and finite-size scaling function $f\left[\xi(n) / L\right]$ converges to a constant value.
In the other limit, for the finite-size systems at the percolation threshold
[i.e. $\xi(n) / L \to \infty$], conductivity has a nonzero value
$\sigma \sim (L/l)^{-t/\nu}$ (Ref.~\cite{2003_ipt_Stauffer}). Therefore, the finite-size
scaling function should have a form $f\left[\xi(n) / L\right] \sim \left[\xi(n) / L\right]^{t /\nu}$,
to cancel the conductivity dependence on density in Eq.~(\ref{sigma_L_l}). Since the
finite-size scaling function $f\left[\xi(n) / L\right]$ above the percolation threshold
is a continuous and smooth function~\cite{2003_ipt_Stauffer}, we approximate it by a
combination of its two limiting behaviors
\begin{equation}
f\left[\frac{\xi(n)}{L}\right] \sim 1 + c (n - n_c)^{-t}(L / l)^{-t / \nu}, \label{f}
\end{equation}
where $c$ is the finite-size parameter. Inserting Eq.~(\ref{f}) into Eq.~(\ref{sigma_L_l}) the
first-order approximation of the finite-size scaling law for conductivity becomes
\begin{equation}
\sigma \sim (n - n_c) ^ t + c (L / l)^{- t / \nu}. \label{sigma_L_l_2}
\end{equation}
Finally, incorporating the finite-size effects given by Eq.~(\ref{sigma_L_l_2}) into the
conductivity model for an infinite-size system, Eq.~(\ref{cond_model_inf}), we obtain
the finite-size model for conductivity
\begin{equation}
\sigma = a \frac{(n - n_c) ^ t + c (L / l)^{- t / \nu}}
{b n^{t - 1} / G_s + (n + n_c)^{t - 2} / G_j}. \label{cond_model_fin}
\end{equation}

\FIGURE{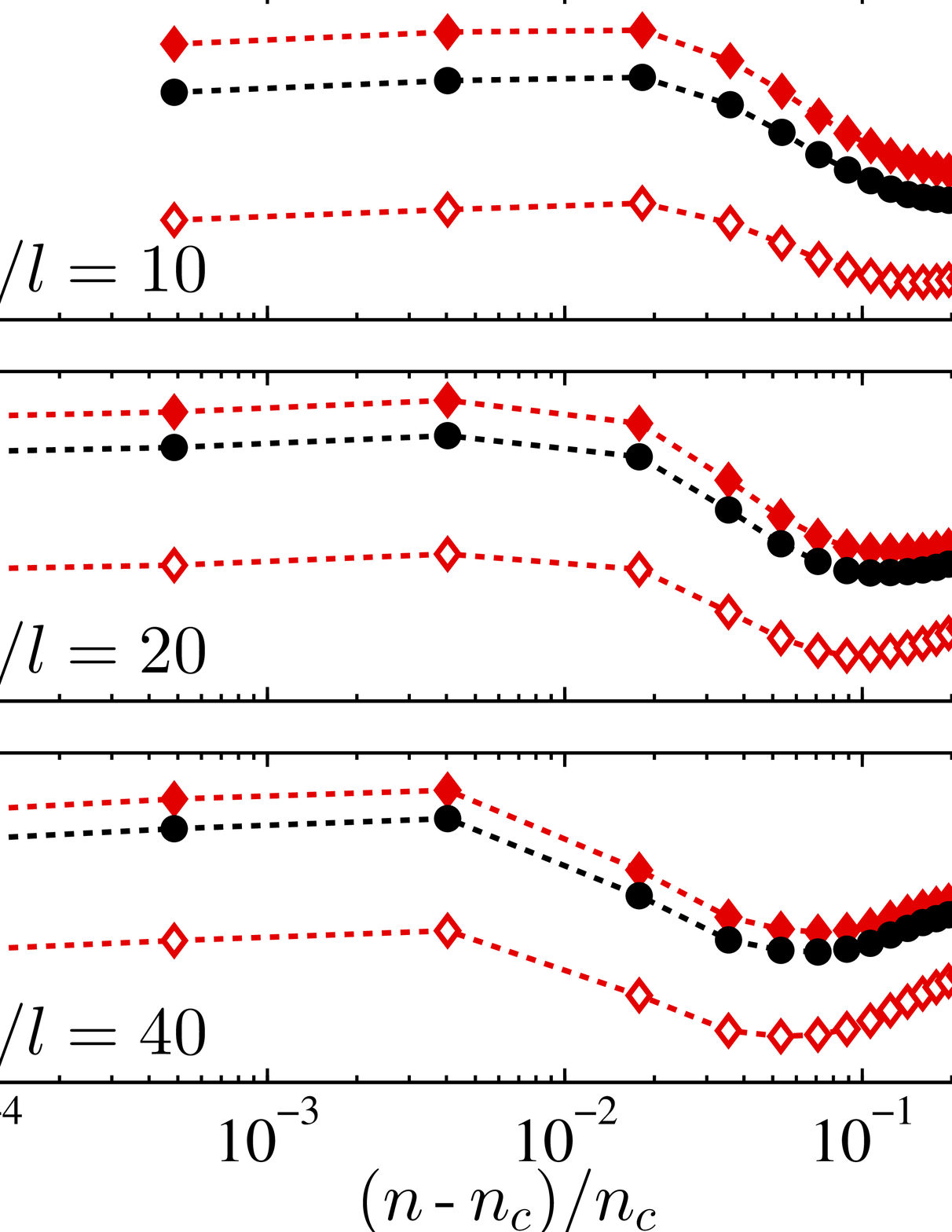}{(Color online) The conductivity ratio between the MC simulation
results $\sigma_{\rm MC}$ and the values obtained from the model $\sigma_{\rm model}$
given by Eq.~(\ref{cond_model_fin}) for different system sizes (a) $L/l = 10$, (b) $20$,
and (c) $40$ and for three conductance ratio values $G_j/G_s = 0.01$, $1$, and $100$.
The error bars are smaller than the size of the points.}

The finite-size parameter for 2D stick systems $c = 2.5(1)$ is calculated using
the least squares fitting methods. A comparison between the MC results and
the values obtained from the model given by Eq.~(\ref{cond_model_fin}) is shown in
Fig.~\ref{figure_4}. The dashed lines in Fig.~\ref{figure_4} denote the local
conductivity exponents $t(n)$ calculated from the model including finite-size
effects, Eq.~(\ref{cond_model_fin}), for systems with increasing size
$L/l = 10$, $20$, and $40$ and for three conductance ratio values $G_j/G_s = 0.01$,
$1$, and $100$. We see that the introduction of finite-size effects in the model
significantly improves the quantitative description of the system close to the
percolation threshold. Finally, the MC conductivity values normalized with the stick
conductance $G_s$ and fitted by Eq.~(\ref{cond_model_fin}) for the systems of size
$L/l = 20$ and conductance ratios from $G_j/G_s=0.001$ to $1000$ are shown in
Fig.~\ref{figure_5}(a). For all studied values of the conductance ratio $G_j/G_s$
the conductivity obtained from the model agrees with the MC results over the whole
range of the stick density $n$, see Fig.~\ref{figure_5}(a). The agreement between
the MC results and the model is good for higher stick densities ($n > 2n_c$)
(i.e., further away from the percolation threshold), but not so good in the
vicinity of the percolation threshold [cf. Fig.~\ref{figure_5}(b)]. Hence, in
the vicinity of the percolation threshold the conductivity ratio between the
MC simulation results and the values obtained from the model is shown
in Fig.~\ref{figure_6} for different system sizes $L/l=10$, $20$, and $40$.
For all three system sizes in Fig.~\ref{figure_6} the curves look qualitatively
similar. Only the density where the dense-system behavior becomes dominant
decreases with the system size $L/l$. To improve the agreement
between the MC results and the model close to the percolation
threshold one could consider a further refinement of the model to include
higher-order correction for the finite-size effect. Finally, the proposed
model for conductivity gives a good estimate of the local conductivity
exponents, as one can see in Figs.~\ref{figure_1} and~\ref{figure_4}.

\section{CONCLUSIONS}
In this paper, we present the results of the numerical Monte Carlo study of the
conductivity of random stick systems for the wide range of densities and junction-to-stick
conductance ratios. We observe the transition from the conductivity of the percolating
cluster to the conductivity of the dense random stick networks with increasing density.
Three limiting cases are identified for the conductivity of whole system: one in the
vicinity of the percolation threshold, and two for high densities when either the junctions
or sticks are superconductive. Each of these cases has a different exponent governing
the power-law dependence of the conductivity from density (i.e., 1.29, 1, and 2,
respectively). As result, the exponent can take values anywhere in the range $(1,2)$
depending on the junction-to-stick conductance ratio. For finite-size systems the
density-dependent exponent can even take values lower than 1. Therefore, it is not appropriate 
to use a simple scaling law to describe the conductivity dependence on the density 
both for finite and dense systems. We instead propose a comprehensive conductivity model, 
derived from the behavior of the limiting cases. We find that the proposed description
gives a satisfactory estimation of the conductivity and the local conductivity exponent
(which is related to the first derivative of the conductivity) over the whole range of the
stick density values. Finite-size effects, important for many practical realizations of the
random conducting networks, are also included in the conductivity model. The presented
methodology could be used to describe the properties of other conducting systems
(i.e., disks, spheres, and fibers).

The authors acknowledge support by the Ministry of Science of the
Republic of Serbia, under Projects No. ON171017 and No. III45018. Numerical
simulations were run on the AEGIS e-Infrastructure, supported in
part by FP7 projects EGI-InSPIRE, PRACE-1IP, PRACE-2IP, and HP-SEE.
The authors also acknowledge support received through SCOPES Grant
No. IZ73Z0--128169 of the Swiss National Science Foundation.



\begin{references}
\bibitem{my_1} For a rectangular system of size $L_x \times L_y$, where
$L_x$ is the length of the electrodes and $L_y$ is the distance between them,
the relation between the system conductance $G$ and conductivity $\sigma$ is
according to Ohm's law given by $G = \sigma L_x / L_y$. In this paper
the system is the square-shaped $L_x = L_y = L$, which implies $G = \sigma$.
Although in the square system conductance is equal to the system conductivity,
for clarity we have assumed terminology in which the {\it conductance} is
used for denoting a single conductive element (stick or junction) while the
{\it conductivity} is related to the entire system.
\end{references}
\end{document}